# Structural Lens Based on Variable Thickness Structures

**Liuxian Zhao[1,2], Chuanxing Bi[1], Miao Yu[2,3,*]**

[1]Institute of Sound and Vibration Research, Hefei University of Technology, 193 Tunxi Road, Hefei 230009, China

[2]Institute for Systems Research, University of Maryland, College Park, MD, 20742, USA

[3]Department of Mechanical Engineering, University of Maryland, College Park, Maryland 20742, USA

*Author to whom correspondence should be addressed: mmyu@umd.edu

Keywords: Triple Focusing; Triple Beam Splitter; Flexural Waves; Plate Structure; Luneburg Lens

**ABSTRACT**

In this article, we report a lens design based on a concentric circular structure with continuous changing of thickness defined in a thin plate structure for focusing a plane wave into three spots (triple focusing) and for splitting elastic waves emanating from a point source into three collimated beams of different directions (three-beam splitting). Inspired by the principle of optical graded index triple focusing lens, the governing equations of the gradient refractive index profiles necessary for achieving such structural lens were obtained. The refractive index profiles were realized by using a lens design with two concentric circular areas of different thickness variation profiles defined in a thin plate. Analytical, numerical, and experimental studies were conducted to investigate the functionalities of the variable thickness structural lens. The results showed that the lens developed in this study were able to perform triple focusing and three-beam splitting with broadband property. Furthermore, the locations of focal points and directions of collimated beams can be engineered by changing the lens thickness profiles according to the governing equations. In addition, the proposed lens is miniature and simple design, which overcome the limitations of previous triple focusing and beam splitters.

## 1. Introduction

In recent years, composite structures and metamaterials have received much attention because of their unique acoustic wave control and manipulation properties. These structures have been shown to lead to enhanced performance in many applications including damage detection [1, 2], beam steering [3-6], energy harvesting [7-9], and signal processing [10, 11], vibration control [12-14], underwater detection [15, 16]. To enhance vibrational energy harvesting [17-20], various elastic wave focusing structures were reported. For example, Qi *et al* [21] showed that a defect region in a square array of silicone rubber cylinders on top of a thin aluminium plate can be used to concentrate elastic wave at the resonant frequency. In another study, Carrara *et al* [22] demonstrated an acoustic mirror consisting of stubs arranged along the outline of an ellipse for focusing reflected waves in the frequency range of 30 – 50 kHz. Furthermore, an acoustic black hole (ABH) realized through thickness variation in a thin plate was shown to achieve elastic wave focusing [23-27]. All of these studies focused on single-location wave focusing.

To increase the number of on-board energy harvesters, multi-location wave focusing techniques were also investigated [28]. Schlichting *et al* [29] proposed passive multi-location schemes by combining multiple wave focusing techniques onto a single platform. However, these schemes increased the size and cost of the multi-source energy harvesting device. More recently, Zhao and Zhou [30] proposed a tunable multi-location wave focusing technique based on frequency selective structures. However, external energy input was needed to realize tuning of the structure frequency.

On the other hand, there is a great need for beam splitters to generate multiple beams from a single source, which has potential applications in beam multiplexing [31], multi-beam structural health monitoring [32], and ultrasonic medical imaging [33]. Phononic crystal structures and metamaterials have been explored for splitting an incident beam into two or more output beams. For example, Lee *et al* [34] designed a beam splitter based on elastic phononic crystal plates for generating self-collimated waves at a single frequency. Jin *et al* [35] proposed a broadband, omnidirectional beam splitter based on graded phononic crystals. These structures require the design of discrete unit cells, which can be complicated.

In this work, we propose a gradient-index (GRIN) design, based on a variable thickness structure, to produce three focal points (*i.e.,* triple focusing) from an input plane flexural wave (Figure 1a), which can benefit multi-source energy harvesting applications. This structural lens can also be used to split flexural waves from a point source into three-beams (*i.e.,* three-beam splitter) (Figure 1b), which can be useful for structural health monitoring, medical imaging, and signal processing applications. Because this lens does not rely on phononic lattices made of discrete unit cells, it offers the advantages of broadband performance compared to previous structures [34-38]. The superiorities of the proposed lens are: (1) the proposed lens is bi-functional for both triple focusing and beam splitter; (2) the proposed lens can achieve triple focusing without increasing the size and cost of the multi-source energy harvesting device; (3) the proposed lens can achieve triple focusing without adding external energy input to realize tuning of the structure frequency; (4) the proposed lens can achieve beam splitter based on the continuous variable thickness structure, which doesnot requires discrete unit cells, therefore, ultra-broadband performance can be achieved.

In the following sections, the governing equations of the variable thickness structural lens will first be provided. Using these equations, triple-focusing lens (three-beam splitters) with different focal lengths and focal spot separation distance (beam splitting angles) will be designed. The performance of the lens will then be investigated through analytical, numerical, and experimental studies.

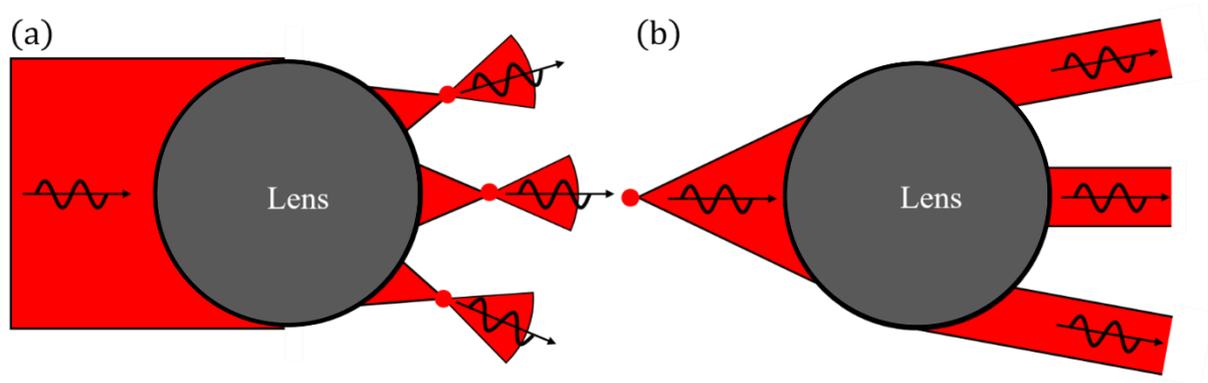

**Figure 1: Schematic of structural lens for (a) triple focusing and (b) three-beam splitting.**

## 2. Structural Lens Design

Luneburg lens is a gradient refractive index lens, whose refractive index varies along the radial direction. The refractive index is achieved through thickness changing of the thin plate structure. This section provides details about the structural lens design. The derivation of refractive index of the proposed lens is also presented.

Luneburg lens is a gradient refractive index lens, which can be used to manipulate optical [39-44] and acoustic waves [45-50]. The original triple-focus lens was developed based on the optical Luneburg lens [51]. The governing equations that relate the locations of the focal points to the refractive index distribution of the triple-focus lens were also derived for electromagnetic waves [52]. The triple-focus lens is essentially a combination of two concentric lenses, each with its own graded refractive index profile that varies radially. The radii of the inner and outer lens are *a* and *R*, respectively (Figure 2(a)). The elastic waves interact with both the inner and outer lenses will be focused onto the central focal point (**P2**), while the waves interacting with the outer lens will be focused onto two symmetric points, **P1** and **P3**, which are separated by a distance of *d* from **P2** (Figure 2 (a)). The distribution of the refractive index (*n*) with the relation of radial distance (*r*) can be expressed as [52]:

$$\begin{cases} n = e^{\left\{\omega(s,F) - \frac{\sin^{-1}\left(\frac{d}{F}\right)}{\pi} \ln \frac{1+\sqrt{1-s^2}}{P_a + \sqrt{P_a^2 - s^2}}\right\}}, & 0 \leq s < P_a \\ n = e^{\left\{\omega(s,F) - \frac{\sin^{-1}\left(\frac{d}{F}\right)}{\pi} \ln \frac{1+\sqrt{1-s^2}}{s}\right\}}, & P_a \leq s \leq R \end{cases}, \quad (1)$$

where $s = rn$, $F$ is the focal length, and $P_a$ is given as:

$$P_a = an(a). \quad (2)$$

Here, $n(a)$ is the distribution of refractive index at the inner circle. The range of $d$ is $0 \leq d \leq d_{max}$, where

$$d_{max} = F\sin\{\pi\omega(P_a, F)/\ln\frac{1+\sqrt{1-P_a^2}}{P_a}\}. \quad (3)$$

In addition, $\omega(s, F)$ can be obtained with the following equations:

$$\begin{cases} \omega(s,F) = \frac{1}{\pi}\int_0^s \frac{\sin^{-1}(k/F)}{\sqrt{k^2-s^2}} dk, & 0 \leq s < P_a \\ \omega(s,F) = \frac{1}{\pi}\int_s^1 \frac{\sin^{-1}(k/F)}{\sqrt{k^2-s^2}} dk, & P_a \leq s < R \end{cases}, \quad (4)$$

where $k$ is a constant for a given ray and $k = nr\sin(\varphi)$, and $\varphi$ refers to the angle between the position vector ***r*** and the tangential ray.

If an excitation point source is placed at the edge of a triple-focus lens, the lens will act as a beam splitter, dividing the source into three-beams on the opposite side of the lens with angles of -*θ*, 0° and +*θ* (Figure 2(b)), where

$$\theta = \sin^{-1}\left(\frac{d}{F}\right). \tag{5}$$

Based on Equation (5), $\theta$ as a function of $d$ is plotted for $F = R$ (red line) and $F=1.5R$ (blue line) in Figure 2(c). It can be observed that when $d = 0$, a zero splitting beam angle $\theta$ is obtained, indicating that there is only one outgoing beam at $\theta = 0°$. For $F = R$ and $d=0.05$ m, a splitting angle of $\theta = 30°$ can be realized ($\theta = 19.5°$ for F=1.5 R and $d=0.05$ m). Note that for $F = R$, the maximum splitting angle of $\theta_{max} = 42.6°$ is reached at $d_{max} = 0.0677$ m. For $F=1.5R$, $\theta_{max}$ (26.2°) is obtained at $d_{max} = 0.0662$ m.

When flexural waves propagate through a thin plate structure with a variable thickness, the phase velocity $c_p$ as a function of its thickness $h$ can be expressed as $c_p = \left(\frac{\omega^2 h^2 E}{12(1-v^2)\rho}\right)^{\frac{1}{4}}$, where $\omega$ is the angular frequency, $\rho$ is the mass density, $E$ is the Young's modulus, and $v$ is the Poisson ratio of the thin homogenous plate. According to Snell's law, $n = \frac{c_0}{c_p}$, where $c_0$ is the phase velocity of flexural wave propagation through a thin plate with a constant thickness $h_0$. The distribution of refractive index is then derived as a function of plate thickness as $n = \sqrt{\frac{h_0}{h}}$ [53]. Combining this refractive index equation with Equation (1), we obtain

$$\begin{cases} h = e^{2\left\{\omega(s,F) - \frac{\sin^{-1}\left(\frac{d}{F}\right)}{\pi} \ln\frac{1+\sqrt{1-s^2}}{P_a+\sqrt{P_a^2-s^2}}\right\}} h_0, & 0 \leq s < P_a \\ h = e^{2\left\{\omega(s,F) - \frac{\sin^{-1}\left(\frac{d}{F}\right)}{\pi} \ln\frac{1+\sqrt{1-s^2}}{s}\right\}} h_0, & P_a \leq s \leq R \end{cases} \tag{6}$$

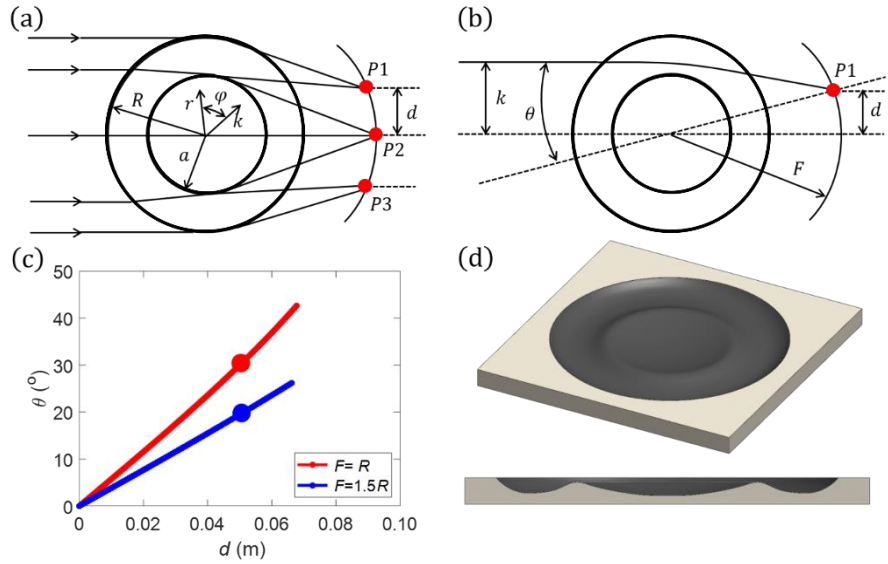

**Figure 2:** Schematic of design principles of the structural lens for (a) triple focusing and (b) 3-beam splitting. (c) Beam angle $\theta$ as a function of the focal point distance $d$, obtained for different focal length $F$. Here, $R = 100$ mm and $P_a = 50$ mm. The values of $a$ are 456 mm and 476 mm for $F=R$ and $F=1.5R$, respectively. (d) Representative design of variable thickness structural lens defined in a thin plate: isometric view (top) and cross-sectional view (bottom).

Based on Equation (6), the thickness profile of the lens can be obtained. A representative design is shown in Figure 2(d). For a case of study, we chose $R = 100$ mm and $h_0 = 4$ mm in the following analyses. Different structural lenses with focal lengths of $F = R$ and $F=1.5R$ were investigated for $d = 0$ m, $d = P_a$, and $d = d_{max}$. The distributions of refractive indices and thicknesses are calculated and provided in Figure 3.

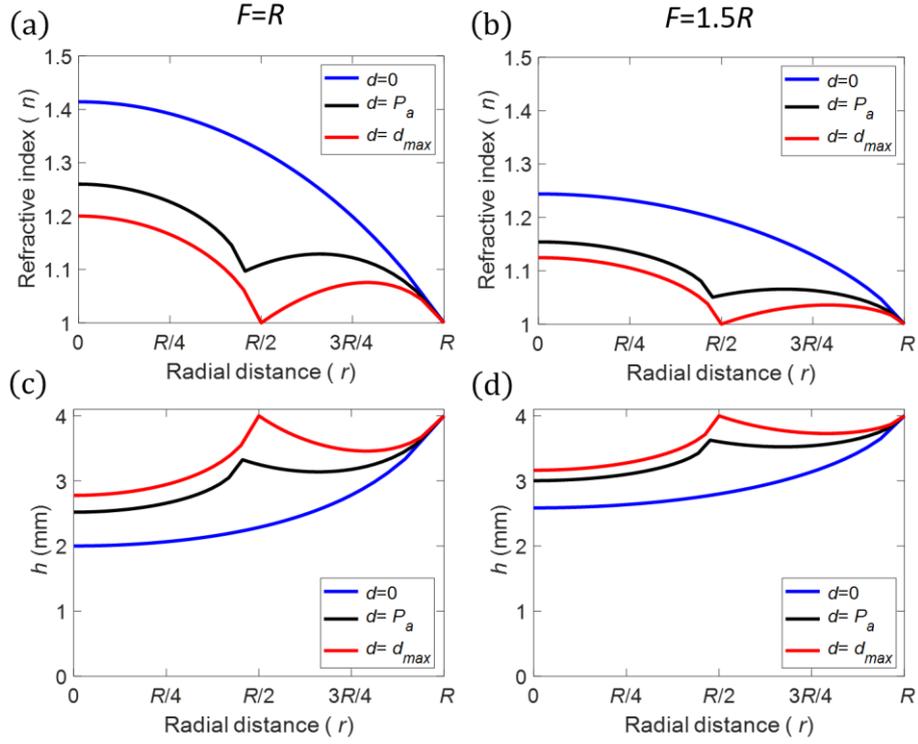

**Figure 3:** Structural lens refractive index with the relation of radial distance for different *d*: **(a)** *F=R* and **(b)** *F=1.5R*. Structural lens thickness variation with the relation of radial distance for different d: **(c)** *F=R* and **(d)** *F=1.5R*. Here, *R* = 100 mm and $P_a$ = 50 mm. The values of *a* are 456 mm and 476 mm for *F=R* and *F=1.5R*, respectively. The values of $d_{max}$ are 677 mm and 662 mm for *F=R* and *F=1.5R*, respectively.

Note that for a propagating ray from a plane wave, the derived refractive index distribution helps ensure that the ray is focused at a particular focusing point on the other side of the lens. Due to the reciprocity of waves, if a ray starts from the focusing point, it will propagate reversely along the same trajectory. In this work, the refractive index distribution of the proposed lens is derived to focus a plane wave into three points, and reversely, transform waves from a point source into three collimated beams. Because the refractive index is independent of the incident frequency, the proposed lens can achieve broadband functionality. This means that the lens works for all the frequencies if the wavelength is smaller than the radius of the lens.

# 3. Results

Numerical and experimental methodologies are described in this section. First, a numerical model of the proposed lens is established and analysed in Section 3.1. Then, the experimental methodology consisting of the fabrication, experimental setup, and results is outlined in Sections 3.2.

## 3.1 Numerical Simulations

Finite element analyses by using COMSOL software were conducted to investigate the triple focusing and three-beam splitting using the designed lens. In the simulation, the dimensions of the structure are 650 mm × 650 mm × 4 mm, the Young's modulus of the structure is 70 GPa, the density is 2700 kg/m$^3$, and the Poisson's ratio $v = 0.33$. Absorbing boundary conditions were used for both time and frequency responses to minimize the reflections from boundaries.

To investigate the broadband triple focusing property for different focal lengths, numerical simulations were performed for $F=R$ and $F=1.5R$ with a designed distance $d = 0.05$ m (red and blue dots in Figure 2(c)). A plane wave located at $x = -150$ mm ($x = 0$, $y = 0$ is the center of the lens) was used for excitation at four different frequencies from 50 kHz to 200 kHz with 50 kHz step. In Figures 4(a)-(b), it can be observed that at each frequency, the simulated triple focusing of flexural waves agrees well with the analytically calculated ray diagrams for both $F=R$ and $F=1.5R$. The distance between the focal points is consistent across all the simulated frequencies for both focal lengths, at around $d = 0.05$ m, which was the designed distance.

For flexural wave beam splitting, a point source at four different frequencies from 50 kHz to 200 kHz with 50 kHz step was generated at the respective focal points of $F=R$ and $F=1.5R$. The simulated beam splitting patterns of flexural waves agree well with the analytically calculated ray diagrams (Figures 4(c)-(d)). The simulation results clearly demonstrated the three-beam splitting characteristics of the structural lens, with the splitter of *F=R* producing a larger splitting angle (31°) than that (22°) produced by the splitter of *F=1.5R*. These splitting angles are in good agreement with the designed angles (30° and 19.5°) for *d* = 0.05 m and remain unchanged for all the frequencies investigated, indicating a good broadband performance of the lens.

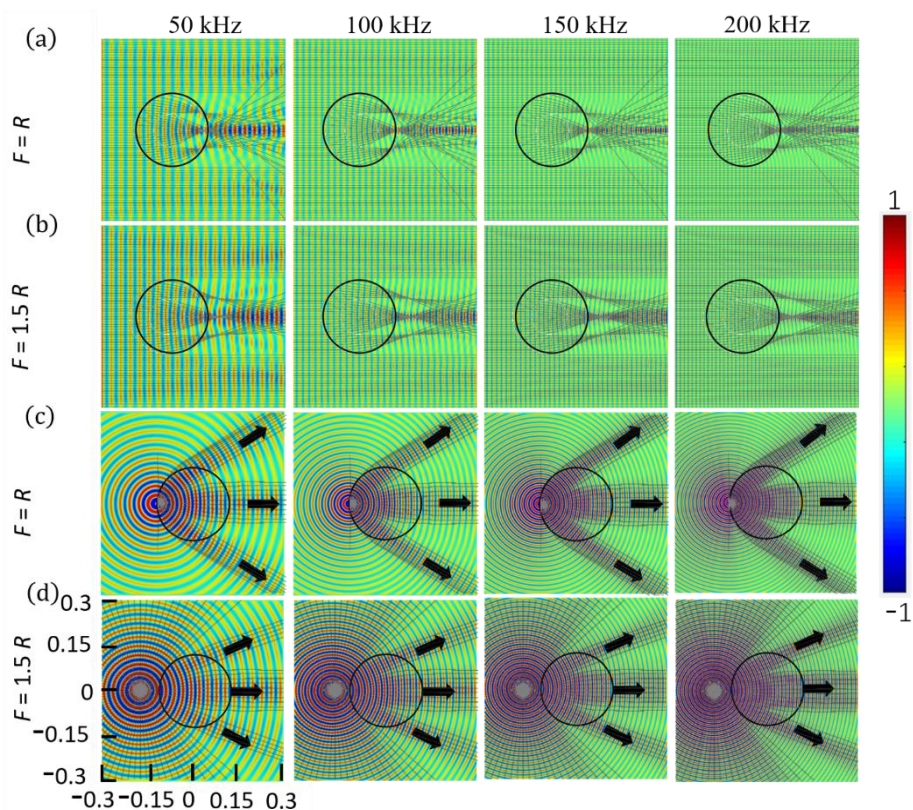

**Figure 4: Finite element simulations of the structural lens for triple focusing under steady state analysis with focal lengths of (a) *F=R* and (b) *F=1.5R*. Note that the ray trajectories are overlaid with the numerical results. Flexural wave beam splitting for focal lengths of (c) *F=R* and (d) *F=1.5R*. The black circle represents the profile of the structural lens. The black arrows represent the beam directions. The colour bar represents the normalized displacement field with respect to the maximum displacement.**

In addition, time-resolved simulations of the triple focusing of flexural waves were also performed for $F=R$ and $F=1.5R$ with a fixed $d = 0.05$ m. For these simulations, a line source along $x$ direction was generated with a central frequency of 100 kHz, which has a frequency range from 50 kHz to 150 kHz, and a 3 dB bandwidth of 67 kHz (from 66 kHz to 133 kHz). The obtained transient responses are shown in Figures 5(a)-(b). At 0.03 ms, the generated plane flexural waves were observed to propagate forward. From 0.07 ms onwards, the structural lens caused the wavefronts to become curved, and the flexural waves were separated into three beams, eventually concentrating in 3 different spots at the designed focal lengths of $F=R$ (Figure 5(a)) and $F=1.5R$ (Figure 5(b)). The focal points for $F=R$ occurred on the edge of the lens, and those for $F=1.5R$ were around $0.65R$ away from the edge of the lens. Across all the simulations, the distance between the focal points agreed well with the designed distance of $d = 0.05$ m.

Flexural wave beam splitting at different time steps was also explored for $F=R$ and $F=1.5R$ with $d = 0.05$ m. A point source excitation was generated at the focal locations ($F=R$ and $F=1.5R$). The time domain responses are shown in Figures 5 (c)-(d). Initially, the generated flexural waves were observed to propagate with a circular wavefront. As the propagating wave interacted with the structural lens, the wavefronts started to flatten and split into three collimated beams. As these three beams exited the lens, they were observed to continue propagating forward with plane wavefronts.

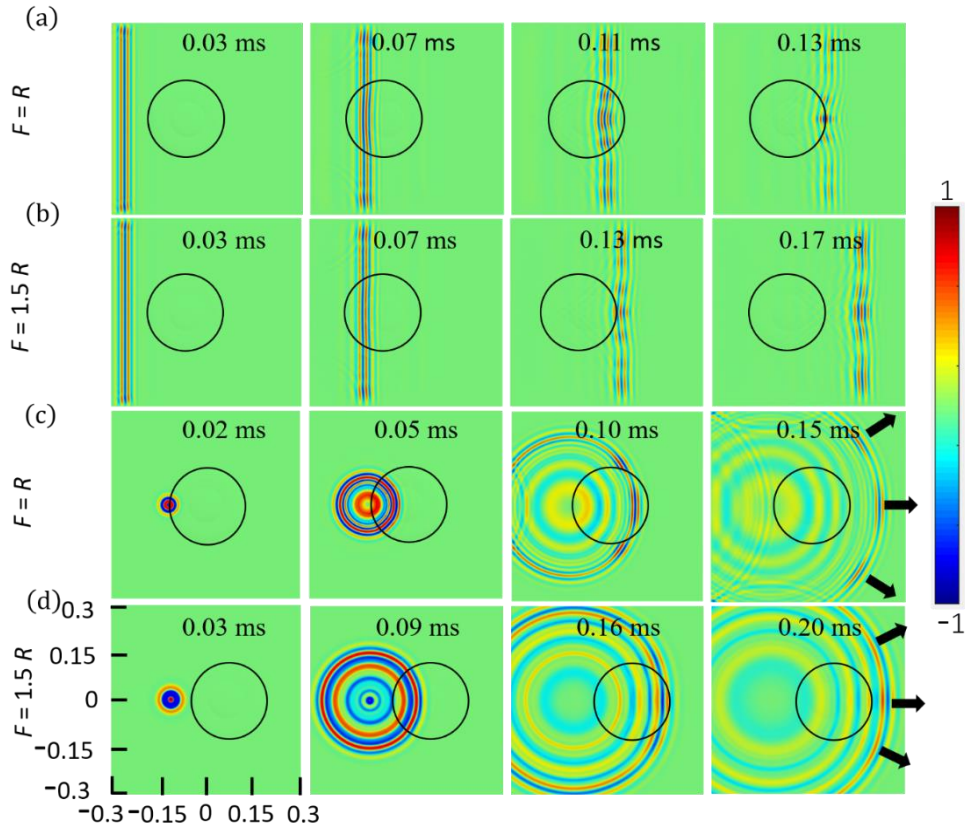

**Figure 5:** Finite element simulations of the structural lens for triple focusing and beam splitting under time domain with focal lengths of (a) *F=R* and (b) *F=1.5R* for different time instances. Three-beam splitting of flexural waves with focal lengths of (c) *F=R* and (d) *F=1.5R* for different time instances. The black circle represents the profile of the structural lens. The black arrows represent the beam directions. The colourbar represents the normalized displacement field with respect to the maximum displacement.

### 3.2 Experimental Results

For experimental validation, the structural lenses were fabricated on thin aluminium alloy 6061 plates (McMaster-Carr). The size of the plate is 650 mm × 650 mm× 4 mm, which are the same as those used in the numerical simulations. Absorbing material was attached to the structural boundaries to minimize reflections from the boundaries. Ten piezo plates (20 mm× 15 mm × 1 mm) were applied to excite a plane wave and a piezo disc (the thickness is 0.6 mm, and the radius is 6 mm) was applied to excite a circular wave (The piezo materials are from STEMiNC Corp). Two lenses with focal lengths of *F=R* (Figure 6 (a)) and *F=1.5R* (Figure 6 (b)) and *d* = 0.05 m were fabricated and tested in the experimental study.

During the experiments, the plate was fixed on a frame with two vertical beams, as shown in Figure 6 (c). A PSV-400 laser vibrometer was applied to measure the displacement data in the out of plane direction through the scan space highlighted in Figures 6 (a) and (b). The measurement was conducted on the structural flat surface at different time instants to verify the performance of triple focusing and beam splitting capabilities using the designed lenses. The frequency of the excitation signal is 100 kHz, and the 3 dB bandwidth of the signal is 67 kHz, which are the same as that used in the numerical simulations. The triple focusing and three-beam splitting of flexural waves were obtained in time domain experimentally and the results are shown in Figure 7. For the triple focusing lens with *F=R*, the focal points occurred around the edge of the lens (Figure 7(a)), and those obtained with the lens of *F=1.5R* were focused outside of the lens (Figure 7(b)). For the beam splitter with *F=R* and *F=1.5R*, the results are provided in Figure 7(c) and (d), respectively. It can be seen that three beams can be generated through a point source.

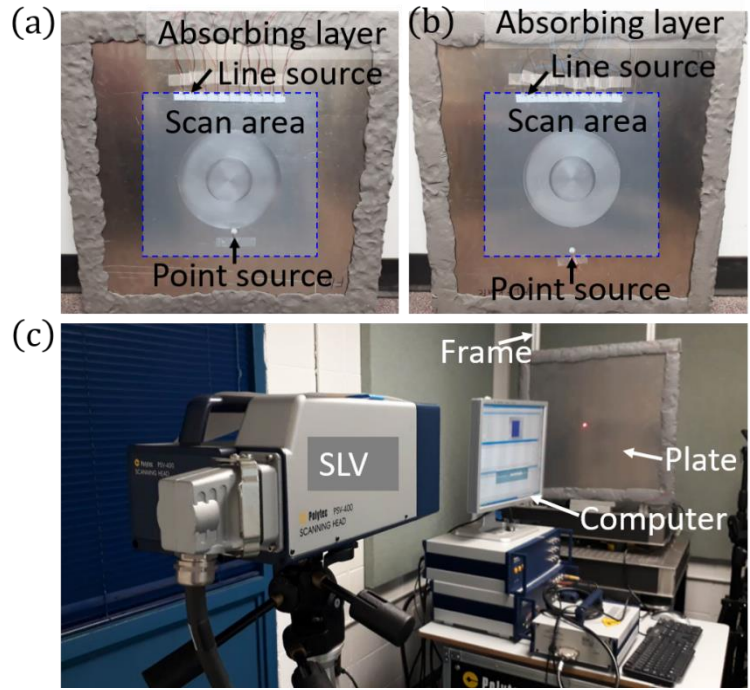

**Figure 6: The fabricated structural lenses with focal lengths of (a) *F=R* and (b) *F=1.5R*. The translucent blue areas indicate the scan areas. (c) Photo of the overall setup. The PSV-400 laser vibrometer was applied to collect the out of plane displacement data in the scan space. The thin plate was covered with clay at its boundaries and constrained by two vertical frames.**

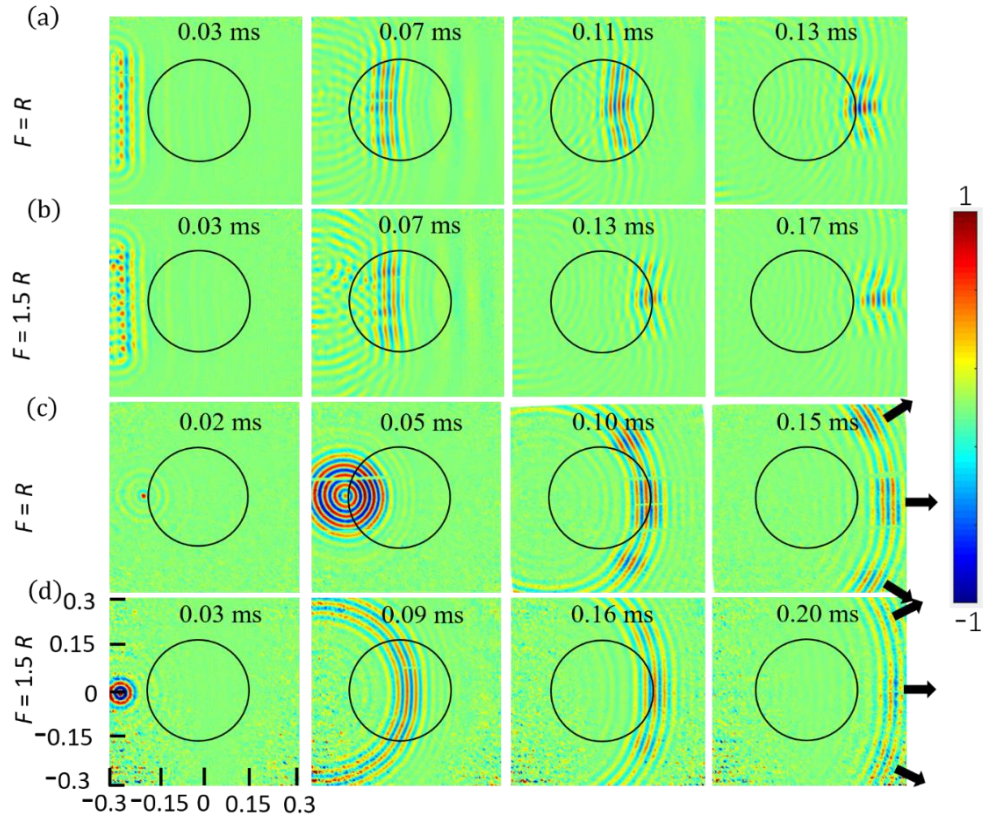

**Figure 7: The performance of both triple-focusing and three-beam splitting using the fabricated variable thickness structural lens in the time domain. The triple focusing of wave with focal lengths of (a) *F=R* and (b) *F=1.5R* at different time instances. Three-beam splitting of flexural waves with focal lengths of (c) *F=R* and (d) *F=1.5R* at different time instances. The black circle represents the profile of the structural lens. The black arrows represent the beam directions. The colour bar represents the normalized displacement field with respect to the maximum displacement.**

## 4. Discussion

The characteristics of the proposed lenses were analysed both numerically and experimentally in the previous sections. In this section, the performance of triple-focusing and three-beam splitting of the proposed lens is discussed.

For beam splitter, it can be seen from Equation (5) that $\theta = 30^o$ and $\theta = 19.5^o$ for $F = R$ and $F=1.5R$, respectively, when $d=0.05$ m. It can be calculated from the numerical simulations that $\theta = 31.2^o$ when $F = R$ and $\theta = 22.5^o$ when $F=1.5R$. These splitting angles are consistent with the calculated values (30° and 19.5°) (Figure 2(c)) for the lens designs with $d = 0.05$ m. Similarly, for three-beam splitting, the measured splitting angles from experimental tests are 30.4° and 26.6° for $F=R$ and $F=1.5R$, respectively. These experimental results validate the three-beam splitting capability of the proposed structural lens.

For triple-focusing, the focal lengths are $F=R$ and $F=1.5R$ for the proposed lenses. The focal points located on the edge of the lens and 0.5R away from the lens, respectively. The distance between the focal points of the designed lens is $d = 0.05$ m. From the numerical simulations, the focal points for $F=R$ occurred on the edge of the lens, and those for $F=1.5R$ were around $0.65R$ away from the edge of the lens. Across all the simulations, the distance between the focal points agreed well with the designed distance of $d = 0.05$ m. From the experimental measurements, for the triple focusing lens with $F=R$, the focal points occurred around the edge of the lens (Figure 7(a)), and those obtained with the lens of $F=1.5R$ were around $0.6R$ away from the edge of the lens (Figure 7(b)). The measured distance between the focal points was close to the designed distance of $d = 0.05$ m. These numerical and experimental results validate the triple focusing capability of the proposed structural lens.

Note that there are slight discrepancies among the analytical results, simulation results, and the experimental measurements, and the triple focusing performance in Figure 8 is not evident. These can be attributed to the following reasons: i) As the theoretical equation of refractive index (Equation (1)) is an approximate equation [52], the designed lens is not perfect. ii) In the analytical equations derived from the ray tracing technique, only ray reflection or refraction were considered, while in the numerical simulations, the structural dimensions, material properties, and different mode shapes were considered. iii) The material properties of the lens used in experiments may not perfectly match with those used in numerical simulations. iv) The adhesive layers used to bond the piezo transducers with the structure in the experimental study were not considered in the numerical simulations, which can generate the inconsistencies between the numerical and experimental results. v) The imperfection of the fabricated structural lens may affect the performance of the lens. vi) The excitations used in the experiments were not a perfect point source or plane wave, while a perfect point source and plane wave excitations were used in the simulations.

## 5. Concluding Remarks

In this study, a novel structural lens was designed, which employs a concentric circular variable thickness structure with two different continuous, gradient refractive index profiles. Analytical, numerical, and experimental studies were performed to demonstrate the performance of the proposed lens. The major findings of this work includes:

- The refractive index is dependent on the thickness of the plate structure, and the relation between the refractive index and the thickness variation of the structure was derived; A novel structural lens was proposed to achieve both triple focusing and three-beam splitting of flexural waves;

- The analytical equations were derived to calculate the refractive index of the proposed lens;

- The refractive index profiles were realized by using a concentric circular variable thickness structure with different thickness profiles;

- Numerical simulations and experimental measurements were conducted to verify the performance of both triple focusing and three-beam splitting of flexural waves, which are consistent with the analytical analysis.

The results demonstrated the lens' ability to focus a plane elastic wave into three spots (triple focusing) and split a point source into three beams with plane wavefront (three-beam splitting). Owing to the simple design and easy fabrication process, this lens holds great promise in energy harvesting, structural health monitoring, medical imaging, and other applications.

**Declaration of Competing Interest**

The authors declare that they have no known competing financial interests or personal relationships that could have appeared to influence the work reported in this paper.

**Data Access Statement**

The data that support the findings of this study are available from the corresponding author upon reasonable request.